\documentclass[reprint,amsmath,amssymb,aps,pra,showkeys]{revtex4-2}

\usepackage{graphicx}
\usepackage{dcolumn}
\usepackage{bm}
\usepackage{comment}
\usepackage{xcolor}

\begin{document}

\title{A Relativistic Tensorial Model \\for Fractional Interaction between Dark Matter and Gravity}

\author{Francesco Benetti$^{1,2,3}$, Andrea Lapi$^{1,2,3,4}$, Samuele Silveravalle$^{1,2,3}$, Stefano Liberati$^{1,2,3}$, Balakrishna S. Haridasu$^{1,2,3}$, Yacer Boumechta$^{1,2,3,5}$, Minahil Adil Butt$^{1,2,3}$, Carlo Baccigalupi$^{1,2,3,6}$}

\address{
\begin{enumerate}
    \item Scuola Internazionale Superiore Studi Avanzati (SISSA), Physics Area, Via Bonomea 265, 34136 Trieste, Italy
    \item Institute for Fundamental Physics of the Universe (IFPU), Via Beirut 2, 34014 Trieste, Italy
    \item Istituto Nazionale Fisica Nucleare (INFN), Sezione di Trieste, Via Valerio 2, 34127 Trieste, Italy
    \item Istituto di Radio-Astronomia (IRA-INAF), Via Gobetti 101, 40129 Bologna, Italy 
    \item The Abdus Salam International Centre for Theoretical Physics (ICTP), Strada Costiera 11, 34151 Trieste, Italy
    \item Istituto Nazionale di Astrofisica (INAF), via Tiepolo 11, I-34143, Trieste, Italy
\end{enumerate}}

\begin{abstract} 
    In a series of recent papers it was shown that several aspects of Dark Matter (DM) phenomenology, such as the velocity profiles of individual dwarfs and spiral galaxies, the scaling relations observed in the latter, and the pressure and density profiles of galaxy clusters, can be explained by assuming the DM component in virialized halos to feel a non-local fractional interaction mediated by gravity. Motivated by the remarkable success of this model, in a recent work we have looked for a general relativistic extension, proposing a theory, dubbed Relativistic Scalar Fractional Gravity or RSFG, in which the trace of the DM stress-energy tensor couples to the scalar curvature via a non-local operator constructed with a fractional power of the d'Alembertian. In this work we construct an extension of that model in which also a non-local coupling between the Ricci tensor and the DM stress energy tensor is present. In the action we encode the normalization between these scalar and tensorial term into two operators $\mathcal{F}_0(\Box)$ and $\mathcal{F}_2(\Box)$, and we derive the general field equations. We then take the weak field limit of the latter, showing that they reduce to general relativity sourced by an effective stress energy tensor, featuring a non local isotropic pressure and anisotropic stress, even if one starts with the assumption of a pressureless DM fluid. Finally, after having worked out the lensing theory in our setup, we test particularly interesting realizations of our framework against the measured convergence profiles of the individual and stacked clusters of the CLASH sample, finding remarkable consistency with the data.
\end{abstract}


\maketitle

%\tableofcontents

\section{Introduction}\label{sec|intro}

A multitude of astrophysical and cosmological probes, including kinematics of spiral galaxies \cite{Rubin_1980,Persic1996},  cosmic microwave background, big bang nucleosynthesis and the baryon acoustic oscillation constraints \cite{Bennet2003,Aver2015,Planck2018,Zhao_2022}, type I$a$ cosmography \cite{Scolnic2018}, X-ray, Sunyaev–Zel’dovich, and strong/weak lensing observations \cite{,Allen2011,Garrel2022,Mantz2022}, reconstruction of mass distribution in the 'Bullet Cluster` \cite{Clowe2006,Paraficz2016}, and others, have firmly established that baryons constitute only some 15\% of the total matter content in the universe, the rest being in the form of a dark matter (DM) component.

The standard cosmological paradigm, dubbed $\rm \Lambda$CDM, envisages DM to be constituted of weakly interacting particles that are non relativistic at the epoch of decoupling, hence they are dubbed 'cold` dark matter or CDM \cite{Bertone_2018}. Since they are weakly interacting with the photon fluid, they do not experience Silk damping, and since they become non relativistic at early times, they do not suffer free-streaming either. As a consequence, bound CDM structures called halos can start to grow before recombination, and after decoupling baryons can fall into the associated gravitational potential wells, rearranging themselves to form galaxies and galaxy clusters. In the present Universe the latter are thus found to be hosted within virialized DM halos, which extend far beyond the edge of visible matter. Remarkably, the density distribution of such halos is predicted from gravity-only $N-$body simulations to follow an approximately universal shape, well described by the Navarro-Frenk-White (NFW) profile $\rho \propto (r/r_s)^{-1}(1+r/r_s)^{-2}$, with $r_s$ being a characteristic scale radius \cite{NFW1996}.

Although on large scales observational data confirm the above picture, in the realm of dwarf galaxies with total masses $\lesssim 10^{11}\, M_\odot$ the situation becomes more uncertain. Galaxy kinematics and/or gravitational lensing data hint to a much flatter density profile in the inner regions (i.e., a core) with respect to the cuspy NFW 
profile (cusp-core problem) \cite{Gentile_2004}; in addition, the observed satellites in Milky Way sized galaxies are found to be much less numerous and less massive than the bound DM halos in $N-$body simulations \cite{Klypin_1999,Boylan_Kolchin_2011} (missing satellites and too-big-to-fail problems); finally, the presence of tight empirical relationships between the properties of the DM and of the baryons in a galaxy, such as the universal core surface density, and the scaling of the core radius with the disk scale length \cite{Salucci_2000,Karukes_2016,Donato_2009,Lapi_2018}, do not find a natural explanation in the standard paradigm. These mismatches between the model and observations represent a severe challenge for the CDM framework at small scales.

Such an observational landscape has motivated the search for alternative scenarios that rely either on non standard DM candidates, such as keV-scale warm DM \cite{Bulbul_2014,Dudas_2014,Ishida_2014,Viel_2005}, very light wave-like particles dubbed fuzzy DM \cite{Hu_2000,shevchuk2023,Schive_2014}, self-interacting DM \cite{Spergel_2000,Loeb_2011,Tulin_2018,Dhanasingham_23}, and others, or on a modification of the gravitational interaction at small acceleration scales like in MONDian frameworks \cite{Milgrom_1983I,Milgrom_1983II}. All these proposals suffer their own issues; for instance, fuzzy DM fails to reproduce some of the aforementioned galactic scaling relations \cite{Hernandez_2023,Burkert_2020}, self interacting DM has been severely constrained by the non-detection of $\gamma$-ray signals in newly discovered dwarf galaxies  \cite{errani2024ursa,crnogorčević2024strong}, and MOND is not able to fully remove the mass discrepancy in galaxy clusters without introducing a DM component, so cheating the original spirit of the theory \cite{Sanders_2003,Hodson_2017}.

In a series of recent works \cite{Benetti_2023, Benetti_2023II,Benetti_2023III} we put forward a fractional gravity (FG) framework which strikes an intermediate course between a modified gravity theory and an exotic DM scenario. It envisages DM in virialized halos to experience a non-local self-interaction mediated by gravity, which can be modeled via fractional operators, featuring derivatives of non-integer order. In a Newtonian setting, the equation governing the DM-gravity interaction is the fractional Poisson equation \cite{GiustiMOND}
\begin{equation}\label{eq|FPE}
    (-\Delta)^s\,\Phi(\textbf{r}) = -4\pi G\,\ell^{2-2s}\rho_{\rm DM}(\textbf{r})
\end{equation}
where $\rho_{\rm DM}(\textbf{r})$ is the DM distribution, $(-\Delta)^s$ is the Fractional Laplacian \cite{fraclap,FLReview}, $s\in[1,3/2]$ controls the strength of the non-local interaction (its range is set to avoid divergences in the theory, with the upper limit being properly defined only in the sense of distributions), while $\ell$ is a length-scale marking the size above which the interaction is strengthened and below which it is reduced with respect to standard Newtonian gravity, recovered for $s=1$.

\begin{figure*}[h]
    \centering
    \includegraphics[width=.75\textwidth]{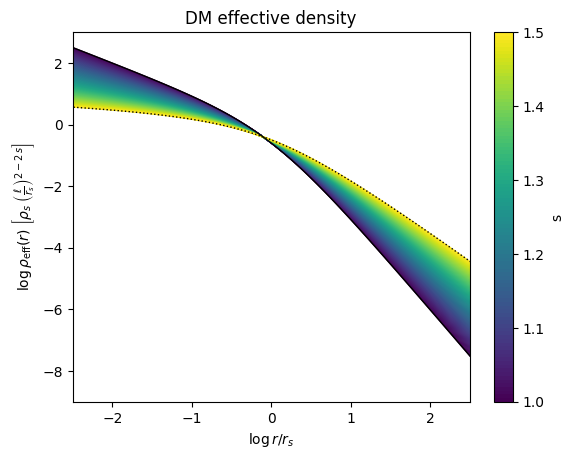}
    \caption{The effective density (normalized to $\rho_s\, (\ell/r_s)^{2-2s}$) vs the radius (normalized to $r_s$), for different values of the fractional index $s$ (color-coded). For reference, the dotted line refers to the maximal value $s = 3/2$.}
    \label{fig.density}
\end{figure*}

If the NFW profile is substituted as an ansatz on the right hand side of Equation \eqref{eq|FPE}, then FG is able to: (i) provide accurate fits to the stacked rotation curves of high and low surface brightness spirals, to the thermodynamic profiles of galaxy clusters, and to the rotation curves of individual dwarf spheroidal and irregular galaxies; (ii) reproduce the observed shape and scatter of the radial acceleration relation (RAR) \cite{McGaugh_2016} over an extended range of galaxy accelerations; (iii) explain the scaling relations observed between the properties of DM halo and the baryonic disk of the host galaxy. All these analyses prove that, when endowed with the fractional interaction, DM performs better that in the Newtonian case, especially in small structures like dwarfs where FG effects are found to be stronger. Remarkably, FG also solves naturally the cusp-core problem, without spoiling the validity of the standard CDM paradigm and the outcomes of gravity-only $N-$body simulations. This is easily seen by realizing that Equation \eqref{eq|FPE} can be reformulated as a classical Poisson equation sourced by a non local effective density 
\begin{equation}\label{eq|DMeff}
    \Delta\Phi(\textbf{r}) = 4\pi G\,\rho_{\rm bar}(\textbf{r}) + 4\pi G\,\rho_{\rm DM,eff}(\textbf{r})
\end{equation}
where we have also added the contribution from baryons, and the two DM distributions are related by $\rho_{\rm DM, eff}(\textbf{r}) = (-\ell^2\,\Delta)^{1-s}\,\rho_{\rm DM}(\textbf{r})$. With $s$
increasing from unity (Newtonian case), the effective density profile corresponding to the NFW distribution progressively flattens in the inner region, where a core-like behavior emerges, while in the outskirts the effective profile resembles an isothermal sphere (see Figure \ref{fig.density}). By looking at the relation between the two distributions in Fourier space, $\tilde{\rho}_{\rm DM,eff}(\textbf{k}) = (\ell\,\lvert\textbf{k}\rvert)^{2-2s}\,\tilde{\rho}_{\rm DM}(\textbf{k})$, one can see that the formation of the core is driven by an energy transfer from larger to smaller modes, i.e. from smaller to larger scales. 

Motivated by the successes of FG at the Newtonian level, in a recent work \cite{Benetti_2024} we proposed a general relativistic extension of the theory. Specifically, we devised an action which naturally extends the Einstein-Hilbert one of general relativity by adding a coupling between the Ricci scalar and the trace of the DM stress energy tensor, mediated by a non local operator $\mathcal{F}(\Box)$, where $\Box = \nabla_\mu\nabla^\mu$ denotes the d'Alembertian. We have derived the field equations for that model, dubbed Relativistic Scalar Fractional Gravity (RSFG for short), showing that it gives rise to a very interesting phenomenology. In particular we were able to: (i) study the weak field limit of the theory, finding that it can be written as standard general relativity sourced by an effective DM stress energy tensor, comprising a density $\bar\rho_{\rm DM}$, an isotropic pressure $\bar p_{\rm DM}$ and an anisotropic stress $\bar\Pi_{\rm DM\,\mu\nu}$; (ii) take the Newtonian limit for a slowly moving source and a nearly static metric, and show that the source $\rho_{\rm DM,eff}$ in Equation \eqref{eq|DMeff} can be written as $\rho_{\rm DM,eff} = \bar\rho_{\rm DM}+3\,\bar p_{\rm DM}$, hinting that the effect of the new interaction can be seen as the action of a non local pressure, which gravitates even in the Newtonian limit and causes the energy transfer from smaller to bigger scales; (iii) derive the post-Newtonian limit to order $c^{-2}$, compute the gravitational $\Phi$ and lensing $\Psi$ potentials, verify that they are different as required by the presence of the anisotropic stress, and check that weak lensing is not modified with respect to general relativity since the sum $\Phi+\Psi$ yield twice the Newtonian potential; (iv) verify that gravitational waves travel at the speed of light, although new massless scalar degrees of freedom propagate; (v) derive the conservation equation for the DM stress energy tensor in the full theory, show that a new non-local force appears, and suggest a physical interpretation of the latter in terms of Mach's principle.

In this work we study another relativistic extension of FG whose action, alongside the aforementioned scalar term $R\,\mathcal{F}_0(\Box)\,T_{\rm DM}$ features also a tensorial coupling of the form $R_{\mu\nu}\,\mathcal{F}_2(\Box)\,T^{\mu\nu}_{\rm DM}$. The motivation for the introduction of such an additional tensorial coupling is manifold: first, together with the scalar term, it provides the most general model that can be constrained by requiring agreement with Equation \eqref{eq|FPE} in the Newtonian limit. In addition, a tensorial term introduces a departure from general relativity in the analysis of gravitational lensing, allowing a direct comparison of the two theories in the realm of galaxy clusters. Moreover, we anticipate that the tensorial term by itself will originate a theory where both the gravitational potential and the lensing convergence are sourced by the same effective density distributions; this can have a value in systems where kinematic and lensing measurements coincide. Note also that if $\mathcal{F}_2(\Box) = -2\mathcal{F}_0(\Box)$ holds, there is the appealing possibility to combine the two terms to form a coupling that involves the Einstein tensor $G_{\mu\nu}\,\mathcal{F}(\Box)\,T^{\mu\nu}_{\rm DM}$. This provides a generalization of a class of theories featuring a non minimal coupling between gravity and DM, that stem from a disformal transformation of the metric \cite{Gandolfi_2021,Gandolfi_2022,Gandolfi_2023}. Let us mention that, in principle one could also add a term involving the Riemann (or Weyl) tensor contracted with two copies of the DM stress energy tensor $R_{\alpha\beta\mu\nu}\,\mathcal{F}_4(\Box)T^{\alpha\mu}_{\rm DM}\,T^{\beta\nu}_{\rm DM}$. However, since this term is quadratic in the matter energy density, it vanishes in the weak field limit, preventing its non local factor $\mathcal{F}_4(\Box)$ to be constrained by comparison with Equation \eqref{eq|FPE}. 

We stress that both the original FG framework and its relativistic extensions are not necessarily meant to subtend a modified gravity theory, but simply the formalization of an emergent non-local behavior of standard DM in virialized environments. In fact, nonlocality
may originate: at the microscopic level from peculiar properties of the DM particles, e.g., some form of quantum entanglement or non-standard interactions mediated by gravity; at the mesoscopic level related to the fluid, coarse-grained description of the DM particles’ collective behavior in a finite volume; at macroscopic scales by the response of a complex (e.g., clumpy and inhomogeneous) DM distribution to the long-range action of gravitational forces against DM and baryonic particles.

Being FG an effective theory (and as such described by an effective action), one should not expect the FG parameters $s$ and $\ell$ to have the same values across all systems and environments. In fact, \cite{Benetti_2023,Benetti_2023II} have tested FG over an extended halo mass range from $10^9–10^{15}\, M_\odot$ by exploiting
stacked rotation curves of spiral galaxies and joint X-ray/Sunyaev-Zel’dovich observations of galaxy clusters, finding that the strength of FG effects depends (at least to a first-order approximation) on halo mass and tend to weaken toward more massive systems. On the one hand, this reinforces the interpretation of FG as an effective, non-universal theory; on the other hand, this implies that FG can substantially alleviate the small-scale issues of the standard $\Lambda$CDM paradigm while retaining its successes on large cosmological scales (e.g., it does not affect background cosmology).

From this point of view, the use of the NFW profile in Equation (\ref{eq|FPE}), expressing the Newtonian limit of relativistic FG, could be viewed as a perturbative correction to the $\Lambda$CDM halo structure as induced from FG effects. In fact, to be fully self-consistent one should run a cosmological $N-$body simulations in a FG setup, and then investigate the resulting halo structure. However, on the one hand the numerical treatment of non-local effects in space and time would be extremely challenging; on the other hand, to a first approximation one should not expect strong deviations from the perturbative approach followed so far.  

The plan of the paper is as follows. In Section \ref{sec|action} we introduce the action of the relativistic FG theory and derive the associated field equations; in Section \ref{sec|weak} we investigate the weak field limit, derive the Newtonian and post-Newtonian approximations showing that Equation \eqref{eq|FPE} is correctly recovered, and compute the gravitational and lensing potentials; in Section \ref{sec|lens} we test particularly interesting realizations of the model by exploiting gravitational lensing data from the CLASH galaxy cluster sample; finally, in Section \ref{sec|conclusions} we summarize our findings and outline future perspectives.

\section{Action and field equations}\label{sec|action}

As stated in the Section \ref{sec|intro}, the model presented in this work is a generalization of the one introduced by \cite{Benetti_2024}, featuring both a scalar $R\,\mathcal{F}_0(\Box)\,T_{\rm DM}$ and a tensorial $R_{\mu\nu}\,\mathcal{F}_2(\Box)\,T^{\mu\nu}_{\rm DM}$ coupling.
The action of the theory thus takes the form
\begin{align}\label{FGaction}
        & S_{\rm RFG}[g,\psi_{\rm bar},\psi_{\rm DM}] = S_{\rm bar}[g,\psi_{\rm bar}] + S_{\rm DM}[g,\psi_{\rm DM}] + S_{\rm EH}[g] \nonumber\\
        & \nonumber \\
        & + \int_{\mathcal{M}}{\rm d}^4x\; \sqrt{-g}\,\left(\,R\, \mathcal{F}_0(\Box)\, T_{\rm DM} + R_{\alpha\beta}\,\mathcal{F}_2(\Box)\, T^{\alpha\beta}_{\rm DM}\,\right)
    \end{align}    
where $S_{\rm bar}[g,\psi_{\rm bar}]$ and $S_{\rm DM}[g,\psi_{\rm DM}]$ denote the matter actions for baryons and DM respectively, $S_{\rm EH}[g]\equiv (2\kappa)^{-1}\,\int_\mathcal{M}\,d^4x\sqrt{-g}\,R$ with $\kappa\equiv 8\pi\,G/c^4$ is the Einstein-Hilbert action, $T^{\alpha\beta}_{\rm DM}$ is the DM stress-energy tensor and $T_{\rm DM}\equiv g_{\alpha\beta}\, T^{\alpha\beta}_{\rm DM}$ the corresponding trace.

The field equations of the theory can then be derived by varying the action expressed by Equation \eqref{FGaction} with respect to the inverse metric. In order to perform the calculation, one has to recall the variation of the Ricci scalar and the Ricci tensor \cite{Calcagni_2021}
\begin{align}
    & \delta R = (R_{\mu\nu} + g_{\mu\nu}\,\Box - \nabla_{\mu}\,\nabla_{\nu})\,\delta g^{\mu\nu}\;, \nonumber \\
    &\nonumber \\
    & \delta R_{\mu\nu} = \nabla^{\alpha}\nabla_{(\mu}    \,\delta g_{\nu)\alpha}-\frac{1}{2}\Box\,\delta g_{\mu\nu} -\frac{1}{2}g^{\alpha\beta}\nabla_{(\mu}\nabla_{\nu)}\,\delta g_{\alpha\beta}\;,
\end{align}
and the variation of the energy momentum tensor for a perfect pressurelss fluid $T^{\alpha\beta}_{\rm DM} = \rho\,u^{\alpha}u^{\beta}$, which provides a good description of collisionless CDM, reported in \cite{haghani2023variation}
\begin{equation}
    \frac{\delta T^{\alpha\beta}_{\rm DM}}{\delta g^{\mu\nu}} = \frac{1}{2}T^{\alpha\beta}_{\rm DM}\,(\,g_{\mu\nu}-u_{\mu}u_{\nu}\,)\;.
\end{equation}
The most involved part of the calculation is the variation of the non local form factor. This can be achieved by recalling the semi-group integral representation of the inverse fractional d'Alembertian \cite{Calcagni_2021} stemming from the definition of the Euler Gamma function $\Gamma(s) \equiv \int_0^\infty\,{\rm d}\tau\,\tau^{s-1}\,e^{-\tau}$, that reads
\begin{equation}
(-\Box)^{-s} = \frac{1}{\Gamma(s)}\,\int^{\infty}_{0}{\rm d}\tau\;\tau^{s-1}\,e^{\tau\Box}\, ,
\end{equation}
together with the Duhamel's formula for the variation of the exponential of an operator $\mathcal{O}$ in terms of the variation of the operator itself \cite{Bauer_2012}
\begin{equation}
\delta(e^{\tau\mathcal{O}})= \int^{\tau}_{0}{\rm d}q\,e^{q\mathcal{O}}\,\delta\mathcal{O}\,e^{(\tau-q)\mathcal{O}}\;.
\end{equation}
In particular, the variation of the inverse d'Alembertian simply follows
\begin{equation}
    \delta(\Box^{-1}) = -\Box^{-1}\,\delta(\Box)\,\Box^{-1}\;.
\end{equation}
Moreover, the action for the variation of the d'Alembertian $\delta(\Box)$ on a scalar and on a $(0,2)$ tensor can be written explicitly as \cite{Conroy_2014}
\begin{align}
    \delta(\Box)\,T & = \delta g^{\alpha\beta}\,\nabla_{\alpha}\nabla_{\beta}\,T + \frac{1}{2}g^{\alpha\beta}\,\nabla_{\lambda}\,T\,\nabla^{\lambda}\,\delta g_{\alpha\beta} + \nonumber \\
    & \nonumber \\
    & - \nabla^{\alpha}\,T\,\nabla^{\beta}\,\delta g_{\alpha\beta}\, ,
\end{align}
\begin{align}
    \delta(\Box)\,T_{\mu\nu} & = -\delta g_{\alpha\beta}\nabla^{\alpha}\nabla^{\beta}\,T_{\mu\nu} - \nabla^{\beta}\delta g_{\alpha\beta}\nabla^{\alpha}\,T_{\mu\nu} + \nonumber \\
    & \nonumber \\
    & +\frac{1}{2}g^{\alpha\beta}\nabla^{\sigma}\,\delta g_{\alpha\beta}\,\nabla_{\sigma}\,T_{\mu\nu} -\nabla^{\beta}\,T^{\alpha}_{(\nu}\nabla_{\mu)}\,\delta g_{\alpha\beta} + \nonumber \\
    & \nonumber \\
    &  -\nabla_{\sigma}\delta g_{\alpha\beta}\,\delta^{\beta}_{(\mu}\,\nabla^{\sigma}\,T^{\alpha}_{\nu)} + \nabla^{\sigma}\delta g_{\alpha\beta}\,\delta^{\beta}_{(\mu}\,\nabla^{\alpha}\,T_{\nu)\sigma} + \nonumber \\
    \nonumber \\
    & - \frac{1}{2}\Box\,\delta g_{\alpha\beta}\,\delta^{\beta}_{(\mu}\,T^{\alpha}_{\nu)} + \frac{1}{2}\nabla^{\alpha}\nabla^{\sigma}\,\delta g_{\alpha\beta}\,\delta^{\beta}_{(\mu}\,T_{\nu)\sigma} + \nonumber \\
    & \nonumber \\
    & -\frac{1}{2}\nabla^{\beta}\nabla_{(\mu}\,\delta g_{\alpha\beta}\,T^{\alpha}_{\nu)}\, .
\end{align}

After a long and tedious calculation that makes abundant use of the above tensor-calculus and operatorial identities, we find the field equations
\begin{align}\label{RSFGEQ}
    & \frac{G_{\mu\nu}}{\kappa} = T_{\rm bar\,\mu\nu} + T_{\rm DM\,\mu\nu} +(T_{\rm DM\,\mu\nu}-g_{\mu\nu}\,T_{\rm DM})\,\mathcal{F}_0\,R+  \nonumber \\
    & \nonumber \\
    & -2(G_{\mu\nu}+g_{\mu\nu}\,\Box-\nabla_{(\mu}\nabla_{\nu)})\,\mathcal{F}_0\,T_{\rm DM} -2\,R\,\frac{\delta\mathcal{F}_0}{\delta g^{\mu\nu}}\,T_{\rm DM} + \nonumber \\
    & \nonumber \\
    & -\Box\,\mathcal{F}_2\,T_{\rm DM\,\mu\nu} + +2\nabla^{\alpha}\nabla_{(\mu}\,\mathcal{F}_2\,T_{\rm DM\,\nu)\alpha} + \nonumber  \\
    &\nonumber \\
    & -g_{\mu\nu}\,\nabla_{\alpha}\nabla_{\beta}\,\mathcal{F}_2\,T^{\alpha\beta}_{\rm DM} + g_{\mu\nu}R_{\alpha\beta}\,\mathcal{F}_2\,T^{\alpha\beta}_{\rm DM}\, -g_{\mu\nu}T^{\alpha\beta}_{\rm DM}\mathcal{F}_2\,R_{\alpha\beta} + \nonumber \\
    \nonumber \\
    & -\frac{T_{\rm DM\,\mu\nu}}{T_{\rm DM}}T^{\alpha\beta}_{\rm DM}\mathcal{F}_2\,R_{\alpha\beta} - 2R_{\alpha\beta}\,\frac{\delta\mathcal{F}_2}{\delta g^{\mu\nu}}\,T^{\alpha\beta}_{\rm DM} \; .
\end{align}
In obtaining Equation \eqref{RSFGEQ} we have discarded a boundary term, which pops out when one integrates by parts in the variation of the action. Following a reasoning similar to the one explained in detail by \cite{Benetti_2024} (see their Section 2), one can reabsorb such a boundary factor in a Gibbons-Hawking-York counter-term in the action given by Equation \eqref{FGaction} \cite{Gybbons-Hawking,York}, without introducing any additional degrees of freedom in the theory.

\section{Weak field limit}\label{sec|weak}

Since the field Equations \eqref{RSFGEQ} are very hard to deal with, in this work we restrict further analysis to the weak field limit. To this purpose, we routinely assume that it is possible to choose a reference frame where the dynamics of the gravitational field is encoded within a small perturbation over the Minkowski background $g_{\mu\nu} = \eta_{\mu\nu} + h_{\mu\nu}$,  with $h_{\mu\nu}\ll 1$. 
This assumption enables to simplify the equations considerably. 

First, having the metric as a small perturbation requires the stress energy tensor to be at least linear in $h$, so that all the terms in Equation \eqref{RSFGEQ} involving the product of the curvature and the stress-energy tensor can be neglected at the linear order. Furthermore, since the non-local operators $\mathcal{F}_{0,2}(\Box)$ always act on an object linear in $h$, at this order they can be replaced with their expression in Minkowski spacetime $\mathcal{F}_{0,2}(\Box_\eta)$, where $\Box_\eta = -\partial^2_0 + \Delta$. In the following we will omit the subscript $\eta$ for simplicity; hereafter anytime a d'Alembertian operator appears, it must be meant as its expression in Minkowksi space. Finally, it is easy to see that in the weak field limit the contracted Bianchi identities imply that the baryons and DM stress energy tensors must satisfy the usual conservation law $\partial_{\mu}\,(T_{\rm bar}^{\mu\nu} + T_{\rm DM}^{\mu\nu})=0$, in terms of flat partial derivatives. Under the assumption that DM is a pressureless fluid, its stress-energy tensor is separately conserved, meaning that  $\partial_{\mu}T_{\rm DM}^{\mu\nu}=0$, and this condition can be reinserted into the field equations to achieve the cancellation of some terms. Taking into account all these simplifications, Equation \eqref{RSFGEQ}, to linear order in $h$, turns into the more manageable form
    \begin{align} \label{WFL}
    \frac{{G}^{(h)}_{\mu\nu}}{\kappa} & = T_{\rm bar\,\mu\nu} + T_{\rm DM\,\mu\nu} - \Box\,\mathcal{F}_2(\Box)\,T_{\rm DM\,\mu\nu} + \nonumber\\
    & \nonumber \\
    & + 2\,(\partial_{\mu}\,\partial_{\nu}-\eta_{\mu\nu}\,\Box)\, \mathcal{F}_0(\Box)\, T_{\rm DM}\; ,
    \end{align}
in terms of the linearized Einstein tensor 
\begin{align}\label{Einsteinh}
    2\,G_{\mu\nu}^{(h)} & = 2\,\partial_{\alpha}\,\partial_{(\mu}\,h^\alpha_{\nu)}-\partial_{\mu}\,\partial_{\nu}\,h-\Box h_{\mu\nu} + \nonumber\\
    & \nonumber \\
    & -\eta_{\mu\nu}\,\partial_{\alpha}\,\partial_{\beta}\,h^{\alpha\beta} + \eta_{\mu\nu}\,\Box h\,.
\end{align}

\subsection{Newtonian and Post-Newtonian limit}\label{|ssec|PN}

If we further assume the metric to be quasi static and the source to move slowly with respect to the speed of light $c$, we can expand Equation \eqref{WFL} in powers of $1/c$ to obtain the Newtonian and post-Newtonian approximations. In this setting it is customary to express the metric perturbation in terms of potentials, constructed from the matter density and velocity, with real coefficients known as post-Newtonian parameters. In a gravity theory, these must be computed from the field equations, and then allow a direct comparison with experimental data (see e.g. \cite{Gravitation} for a discussion on this procedure, known as Parametrized post-Newtonian, or PPN in short, formalism). If we retain only terms up to order $1/c^{2}$, the metric perturbation in the standard PPN coordinate system can be written as
\begin{equation}\label{standard}
    h_{\mu\nu} = 
    \begin{pmatrix} 
        -\frac{2\Phi}{c^2} &  \\
        & -\frac{2\Psi}{c^2}\,\delta_{ij}
    \end{pmatrix}\; .
\end{equation}
Here $\Phi$ encodes the time-time curvature and corresponds to the potential satisfying the analogous of the Poisson equation in the Newtonian limit of a given gravity theory. On the other hand, $\Psi$ represents the space-space curvature and is usually written as $\Psi = \gamma\,\Phi$ in terms of the PPN parameter $\gamma$, representing the amount of space curvature per unit rest mass. 

Inserting Equation \eqref{standard} into \eqref{Einsteinh} we find, up to order $1/c^2$, the expansion:
\begin{equation}
    G_{00}^{(h)} = \frac{2\,\Delta\Psi}{c^2}, \,\,\,\,\, G_{0i}^{(h)} = 0, \,\,\,\,\, G_{ij}^{(h)} = (\partial_{i}\,\partial_{j}-\delta_{ij}\,\Delta)\,\frac{\Psi-\Phi}{c^2}\;.
\end{equation} 
By performing the expansion at the same order of the right hand side of Equation \eqref{WFL}, and taking into account that for a Newtonian source the spatial stresses $T_{0i}\,,T_{ij}$ are suppressed with respect to the rest mass density $T_{00} = c^2\rho$, we obtain the equations satisfied by the two potentials
\begin{align}
     & \Delta\Phi = 4\pi G\,\rho_{\rm bar} + 4\pi G\,(1 + (2\mathcal{F}_0-\mathcal{F}_2)\,\Delta)\,\rho_{\rm DM}\,,\label{NL}\\
     & \nonumber\\
     & \Delta\Psi = 4\pi G\,\rho_{\rm bar} + 4\pi G\,(1- (2\mathcal{F}_0 + \mathcal{F}_2)\,\Delta)\,\rho_{\rm DM} \label{gamma}\,.
\end{align}

By comparison of Equation \eqref{NL} with Equation \eqref{eq|DMeff}, we find that, for the Newtonian limit of our general theory to coincide with the FG framework adopted in our previous works, the relation
\begin{equation}\label{eq|relation}
    \mathcal{F}_2(\Box)\,-2 \,\mathcal{F}_0(\Box) \,=\,\mathcal{F}(\Box)\,\equiv\,\frac{1-(-\ell^2\,\Box)^{1-s}}{\Box}
\end{equation}
must hold. Since this constraint does not fix the form of the operators uniquely, one can construct different theories by appropriately choosing $\mathcal{F}_0$ and $\mathcal{F}_2$. For instance, the scalar model (relativistic scalar fractional gravity or RSFG) explored in \cite{Benetti_2024} is obtained setting $\mathcal{F}_0=-\mathcal{F}/2$, $\mathcal{F}_2=0$, while a pure tensorial model (relativistic tensorial fractional gravity or RTFG) can be obtained with $\mathcal{F}_0=0$, $\mathcal{F}_2=\mathcal{F}$. Of particular interest is the choice $\mathcal{F}_0=-\mathcal{F}/4$, $\mathcal{F}_2=\mathcal{F}/2$, for which the scalar and tensorial part in the action \eqref{FGaction} combine, leading to a coupling of the type $G_{\mu\nu}\,\mathcal{F}(\Box)\,T^{\mu\nu}_{\rm DM}$ involving the Einstein tensor; we will refer to this latter theory as Einstein Fractional Gravity or EFG for short. 

In passing we note that this form of the coupling resembles a model studied by \cite{Gandolfi_2021,Gandolfi_2022}, called Non Minimal Coupled Dark Matter (NMCDM), in which the non-standard interaction between gravity and DM emerges dynamically from the collective behavior of the coarse-grained DM field (e.g., via Bose-Einstein condensation) with an averaging/coherence length $L$. The NMCDM action, and the field equations following from it, can be obtained by those of EFG by simply setting $\mathcal{F}\equiv \epsilon L^2$, where $\epsilon = \pm 1$ is a parameter of the model. Even if NMCDM is a local theory, as opposed to FG, which is inherently nonlocal, the similarity of the couplings could help to shed light on the physical mechanism behind the latter, an issue which we plan to address in future works.

As we have already shown in \cite{Benetti_2024}, the right hand side of Equation \eqref{WFL} involving DM can be interpreted as a modified stress-energy tensor with both pressure and
anisotropic stress
\begin{align}\label{eq|DMeset}
    T^{\mu\nu}_{\rm DM\,eff} = (\bar\rho_{\rm DM} + \bar p_{\rm DM})\,u^{\mu}u^{\nu} + \bar p_{\rm DM}\,\eta^{\mu\nu} + \Pi^{\mu\nu}\;.
\end{align}
In a frame comoving with the fluid, where $u^{\mu} = \delta^{\mu}_0$, the various components can be found by comparison of Equations \eqref{WFL} and \eqref{eq|DMeset}, requiring that $T_{\rm DM\,eff} = -\bar\rho_{\rm DM} + 3\bar p_{\rm DM}$,
and that $T^{00}_{\rm DM\,eff} = \bar\rho_{\rm DM}$, $\Pi^{00} = 0$. The result reads
\begin{align}
   & \bar\rho_{\rm DM} = \rho_{\rm DM} + (2\mathcal{F}_0 + \mathcal{F})\partial^2_0\,\rho_{\rm DM}-(4\mathcal{F}_0 + \mathcal{F})\Delta\,\rho_{\rm DM}\,,\nonumber \\
   & \bar p_{\rm DM} = 2\left(\,\frac{2}{3}\,\Delta - \partial^2_0\, \right)\,\mathcal{F}_0\,\rho_{\rm DM}\,, \nonumber \\
   & \Pi_{\mu\nu} = 2\left[\left(\partial_{0}^2+\frac{1}{3}\Delta\right)u_\mu u_\nu -\left(\partial_{\mu}\partial_{\nu}-\frac{1}{3}\eta_{\mu\nu}\Delta\right)\right]\mathcal{F}_0\,\rho_{\rm DM}\,.
\end{align}
Is it clear from these expressions that the pressure and the anisotropic stress are present only when $\mathcal{F}_0\neq 0$, i.e. when the action includes a scalar term.
Futhermore the expected relation
\begin{equation}
    \Delta\,(\Psi-\Phi) = 12\pi\, G\,\left(\frac{\partial_{i}\,\partial_{j}}{\Delta}-\frac{1}{3}\,\delta_{ij}\right)\,\Pi^{ij}
\end{equation}
between the difference of the lensing and the Newtonian  potentials and the anisotropic stress holds. It is instructive to look at the combination $\bar\rho_{\rm DM} + 3\,\bar p_{\rm DM}$ for a static source
\begin{equation}\label{eq|d+3p}
    \bar\rho_{\rm DM} + 3\,\bar p_{\rm DM} = \rho_{\rm DM} -\Delta\,\mathcal{F}\,\rho_{\rm DM} = \rho_{\rm DM, eff}\;.
\end{equation}
This expression shows that the effective density sourcing the gravitational potential in the fractional Poisson Equation \eqref{eq|FPE} results from the action of a non local isotropic pressure, which gravitates even in the Newtonian limit. Note that this result is independent on the particular choice of the operators $\mathcal{F}_0$ and $\mathcal{F}_2$, and thus on the particular flavor of the relativistic FG theory.

\section{Testing FG with lensing data of galaxy clusters}\label{sec|lens}

We now aim to test FG by exploiting gravitational lensing data of galaxy clusters. In this context, it is customary to define the angular distances between the source and the observer $D_{s}$, the lens and the observer $D_{l}$, and the source and the lens $D_{ls}$, where in a flat FRW Universe
\begin{equation}
    D_{ij} = \frac{c}{1+z_j}\,\int_{z_i}^{z_j}\,\frac{{\rm d}z}{H(z)}
\end{equation}
holds in terms of the Hubble parameter $H(z) = H_0^2\,[\Omega_{\rm m}\,(1+z)^3 + 1-\Omega_{\rm m}]$. Hereafter the angular distances are computed in the standard $\Lambda$CDM cosmology with parameters from \cite{Planck2018}.

In the thin-lens approximation the lens is considered to lay on a two-dimensional plane, since its dimension along the line of sight is much smaller than the other distances involved. In such a case the lens primary function is to deflect the light beams coming from the source by an angle on the lens plane given by
\begin{equation}\label{angle}
    \hat{\vec\alpha} = \int_{-\infty}^{+\infty}{\rm d}w\;\vec\nabla_{\perp}\,\left(\frac{\Phi(R,w)+\Psi(R,w)}{c^2}\right)\; ,
\end{equation}
where $\vec\nabla_{\perp}$ represents the two-dimensional gradient operator, which is perpendicular to the path of the light, $R$ denotes the two-dimensional projected radius on the lens plane, and $w$ indicates the coordinate that specifies the
position along the line of sight. The deflection angle $\hat{\vec\alpha}$ can be described using the effective lensing potential $\hat{\vec\alpha} = (D_{l}/D_{ls}\,D_{s})\,\vec\nabla_{\perp}\Phi_{\rm lens}$, where
\begin{equation}\label{LensPot}
    \Phi_{\rm lens}(R) = \frac{D_{ls}\,D_{s}}{D_{l}}\,\int_{-\infty}^{+\infty}{\rm d}w\;\left(\frac{\Phi(R,w)+\Psi(R,w)}{c^2}\right)\; .
\end{equation}
The Laplacian $\Delta_{r}$ of Equation \eqref{LensPot}, taken with respect to the three dimensional radius $r = \sqrt{R^2+w^2}$, defines the so called `convergence' 
\begin{equation}\label{conv}
    \kappa(R) = \frac{D_{l}\,D_{ls}}{D_{s}}\,\int_{-\infty}^{+\infty}{\rm d}w\;\Delta_{r}\left(\frac{\Phi(R,w)+\Psi(R,w)}{2\,c^2}\right)\;.
\end{equation}

Neglecting the contribution of baryons, in FG we have from Equations \eqref{NL} and \eqref{gamma} the relevant expressions
\begin{align}
    & \Delta_r\Phi = 4\pi\, G\, (1-\mathcal{F}\Delta)\rho_{\rm DM} = 4\pi\,G\,\rho_{\rm DM, eff}\,, \\
    & \nonumber \\
    & \Delta_r\Psi = 4\pi\,G\,(1-(4\mathcal{F}_0+\mathcal{F})\,\Delta)\rho_{\rm DM}\,,
\end{align}
which yield the convergence
\begin{equation}\label{eq|convergence}
        \kappa(R) = \frac{1}{\Sigma_{\rm crit}}\,\int_{-\infty}^{+\infty}{\rm d}w\;(\rho_{\rm DM, eff}-2\,\mathcal{F}_0\Delta\,\rho_{\rm DM})\;
\end{equation}
in terms of the so called critical surface density $\Sigma_{\rm crit} \equiv (c^2/4\pi G)\,(D_{s}/D_{ls}\,D_{l})$. As already mentioned in the Introduction, \cite{Benetti_2023,Benetti_2023II} have shown that FG effects tend to become negligible on large cosmological scales. This implies that background cosmology and thus the angular distances $D_{ij}$ entering the lensing analysis are not appreciably altered with respect to their standard expressions in $\Lambda$CDM. Moreover, the standard lensing formalism recalled above can be applied since on large scales the metric in the weak field limit over an expanding background can be approximately put in the same form as general relativity \cite{Bartelmann2001}, apart from the difference in the gravitational and lensing potentials induced by FG effects within virialized structures.

In RSFG $\mathcal{F}_0= -\mathcal{F}/2$ holds and the convergence from Equation \eqref{eq|convergence}
\begin{equation}\label{eq|convscalmodel}
\kappa_{\rm RSFG}(R) = \frac{1}{\Sigma_{\rm crit}}\,\int_{-\infty}^{+\infty}{\rm d}w\;\rho_{\rm DM}\; 
\end{equation}
coincides with that of general relativity (see \cite{Benetti_2024}); in this case the gravitational potential is sourced by the effective density $\rho_{\rm DM, eff}$ while the convergence is determined by the true DM density $\rho_{\rm DM}$. In RTFG $\mathcal{F}_0=0$ holds and the convergence reads:
\begin{equation}\label{eq|convtensmodel}
\kappa_{\rm RTFG}(R) = \frac{1}{\Sigma_{\rm crit}}\,\int_{-\infty}^{+\infty}{\rm d}w\;\rho_{\rm DM, eff}\; ;
\end{equation}
in this case both the gravitational potential and the lensing convergence are generated by the same effective DM density $\rho_{\rm DM,eff}$.
In EFG $\mathcal{F}_0 = -\mathcal{F}/4$ holds and the convergence 
\begin{equation}\label{eq|convmodel}
        \kappa_{\rm EFG}(R) = \frac{1}{\Sigma_{\rm crit}}\,\int_{-\infty}^{+\infty}{\rm d}w\;\left(\frac{\rho_{\rm DM, eff} + \rho_{\rm DM}}{2}\right)\;,
\end{equation}
involves a sort of arithmetic average between the Newtonian and the FG density distributions, which retains some difference from the source of the gravitational potential $\rho_{\rm DM, eff}$.

As anticipated in Section \ref{sec|intro} the EFG model can be motivated by its resemblance with the NMCDM model discussed previously, and by the fact that it is the most natural way to combine the scalar and tensorial parts of the action in a single term. However, it may suffer of problems at the scales of galaxies, since recent data (e.g., \cite{Brouwer2021, Mistele2024}) indicate a remarkable agreement among DM mass estimates based from weak lensing (that depend on lensing convergence) and from kinematical measurements of galaxy rotation curves (that depend on the gravitational potential); the tension can be even stronger for the RSFG model and other flavors of the theory. From this point of view, the RTFG model may instead be preferred since it is the only FG framework where such an agreement is ensured by construction. We will therefore present the analysis of cluster lensing data for both the EFG and the RTFG model (recall that as for lensing the RSFG coincides with general relativity).

Specifically, we test FG by exploiting the lensing convergence profiles of the individual and stacked clusters from the CLASH sample \cite{Umetsu_2016,Bouche2022,zamani_2024}. This consists of $20$ galaxy clusters with redshifts in the range $0.2\lesssim z\lesssim 0.7$ for which radial convergence profiles have been reconstructed from a joint analysis of weak lensing shear/magnification data and strong lensing observations. We discard one of the X-ray selected cluster (RXJ1532) from the CLASH sample since no multiple images have been identified in it and therefore the mass reconstruction is based only on wide-field data. For the sake of simplicity, as in \cite{Umetsu_2016,Bouche2022} we report and analyze the CLASH data in terms of $\kappa_\infty(R)$ profiles, meaning that the $\Sigma_{\rm crit}$ is computed in the limit of a very distant source (i.e., $z_s\rightarrow \infty$).

We adopt a Bayesian framework, characterized by the parameter set $\theta\equiv (s,\log \ell/r_s, \log M_{200}, \log c_{200})$, where $s$ and $\ell$ are the FG index and lenght-scale, while $M_{200}$ and $c_{200}$ are the DM mass and concentration at $R_{200}$, the reference radius within which the average DM density is $200$ times the critical density of the Universe, and $r_s\equiv R_{200}/c_{200}$ is the NFW scale radius. We assume a Gaussian log-likelihood
\begin{equation}\label{eq|likelihood}
\ln \mathcal{L}(\theta) = -\chi^2(\theta)/2~,
\end{equation}
where the chi-square $\chi^2(\theta)=\sum_{ij}\,[\kappa_\infty^{\mathcal{M}}(\theta,r_i)-\mathcal{\kappa_\infty^{\mathcal{D}}}(r_i)]\,\mathcal{C}_{ij}^{-1}\, [\kappa_\infty^{\mathcal{M}}(\theta,r_j)-\mathcal{\kappa_\infty^{\mathcal{D}}}(r_j)]$ is obtained by comparing our model expectations $\kappa_\infty^{\mathcal{M}}(\theta,r_i)$ to the data values $\kappa_\infty^{\mathcal{D}}(r_i)$, summing over radial coordinates $r_i$ and taking into account the variance-covariance matrix $\mathcal{C}_{ij}$ among radial bins (see \cite{Umetsu_2016}). More in detail, our fitting model is constituted by $\kappa_\infty(R)$ profiles computed according to Equation (\ref{eq|convtensmodel}) for RTFG or (\ref{eq|convmodel}) for EFG; note that in our analysis we neglect the contribution to the convergence profile from baryonic matter (e.g., stars, hot gas) since this is expected to be subdominant in the range of radii probed by the CLASH data.

We adopt flat priors $\pi(\theta)$ on the parameters $s\in [1,3/2]$, $\log \ell/r_s\in [-3,1]$, $\log c_{200}\in [0,2]$ and $\log M_{200}\, [M_\odot]\in [12,16]$. Moreover, we include a log-normal prior to take into account the relation $c_{200}-M_{200}$ as measured from $N-$body simulations in the $\Lambda$CDM cosmology \cite{Dutton2014,Wang2020} with dispersion around $0.15$ dex. 
Two caveats are in order here. First, we prefer to fit for the ratio $\ell/r_s$ instead of $\ell$ since from the analysis by \cite{Benetti_2023,Benetti_2023II} the latter quantity is not expected to be the same at all scales, but to depend on halo mass and possibly on other environmental properties; specifically, in \cite{Benetti_2023II} we have found that to a first-order approximation it scales with mass very similarly to the NFW scale radius $r_s$, so the ratio $\ell/r_s$ could be more uniform across different systems. However, from \cite{Benetti_2023II} one can anticipate that even the ratio $\ell/r_s$ will be poorly constrained when focusing on massive galaxy clusters, since at these scales the index $s$ is expected to deviate mildly from one (FG effects are expected to be weak), and the overall normalization of the effective mass profile solely depends on $(\ell/r_s)^{2-2s}$. The second caveat concerns the adopted prior on the $c_{200}-M_{200}$ relationship. From the thermodynamic analysis of the XCOP sample \cite{Benetti_2023II} have found that in FG the concentration scales with halo mass very similarly to the $\Lambda$CDM case; this is again a manifestation of the fact that FG effects in massive clusters tend to be weak. Thus in the present analysis we impose a standard $\Lambda$CDM prior on the halo concentration vs. mass relationship; however, we have found a posteriori that the impact of the adopted prior on the outcomes of our analysis is mild, though it helps somewhat in constraining the fractional index $s$ and hence to highlight the fractional deviations from Newtonian gravity.

We sample the parameter posterior distributions $\mathcal{P}(\theta) \propto \mathcal{L}(\theta)\,\pi(\theta)$ via the MCMC Python package \texttt{emcee} \cite{emcee}, running it with $10^4$ steps and $100$ walkers; each walker is initialized with a random position extracted from the priors discussed above. To speed up convergence, we adopt a mixture of differential evolution and snooker moves of the walkers, in proportion of $0.8$ and $0.2$ respectively, that emulates a parallel tempering algorithm. After checking the auto-correlation time, we remove the first $20\%$ of the flattened chain to ensure burn-in; the typical acceptance fractions of the various runs are around $30\%$.

\begin{figure*}[h]
    \centering
    \includegraphics[width=1.\textwidth]{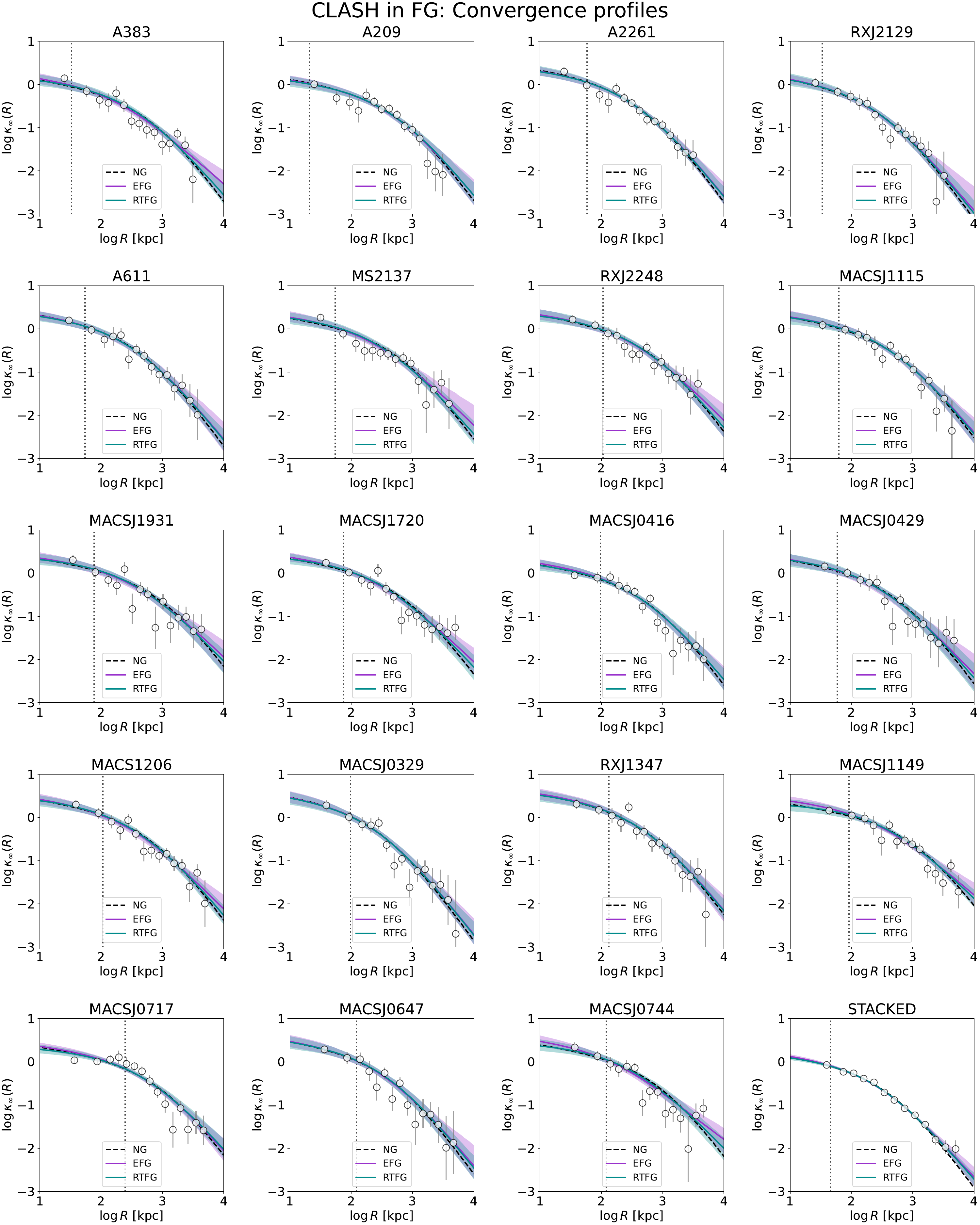}
    \caption{Fits to the lensing convergence profiles $\kappa_\infty(R)$ of the individual and of the stacked CLASH cluster sample according to the EFG (magenta) and RTFG (cyan) model (see Section \ref{sec|lens} for details). Solid lines with shaded areas illustrate the median and the $2\sigma$ credible interval from sampling the posterior distribution. Dashed black lines display, for comparison, the bestfits in general relativity. The vertical dotted line marks the Einstein radius in the lens plane, on assuming a reference source redshift $z_s\sim 2$.}
    \label{fig|convergence}
\end{figure*}

The results of the fit to the convergence profiles $\kappa_\infty(R)$ are shown in Figures \ref{fig|convergence}, where the solid lines with shaded areas illustrate the median and the $2\sigma$ credible interval, in magenta for EFG and in cyan for RTFG; for reference, dashed lines display the bestfits in Newtonian gravity. All in all, both the EFG and the RTFG fits are always very good, both in the inner and outer portions of the convergence profiles. Note, however, that in the outermost radial range the two-halo term (from large-scale clustering) can somewhat alter the profile, and this has not been included in the present analysis.

In Table \ref{tab|fitparams} the marginalized posterior estimates (median and $1\sigma$ confidence intervals) of the parameters are reported. In about half of the clusters 
a deviation of the index $s$ from the Newtonian gravity value of $1$ is favored. The $\ell/r_s$ ratio, due to its degeneracy with the $s$ and the mass $M_{200}$, is marginally constrained, and in many instances only a loose lower limit can be inferred. The estimates of the mass (concentration) tend to be slightly smaller (higher) than in general relativity, though consistent within the $2\sigma$ uncertainties. Unfortunately, the present quality and radial extent of the data does not allow to statistically prefer the FG fits over the general relativity ones, or to discern a clear preference for EFG vs. RTFG, in terms of reduced $\chi^2$ (nor Bayesian evidence criterions). However, we can fairly conclude that both the FG models perform decently well on this data sample, at a level comparable with the standard general relativistic setting.

\section{Summary and outlooks}\label{sec|conclusions}

In this work we have looked for an extension of the scalar relativistic theory developed in \cite{Benetti_2024}, featuring a non local coupling between the Ricci scalar and the trace of the DM stress energy tensor, $R\,\mathcal{F}_0(\Box)\,T_{\rm DM}$, adding a tensorial non local coupling $R_{\mu\nu}\,\mathcal{F}_2(\Box)\,T^{\mu\nu}_{\rm DM}$ between the Ricci and the DM stress energy tensor.

Our main results can be summarized as follows:

\begin{itemize}

\item we have derived the most general field equations starting from an action principle;

\item we have investigated the weak field limit of the theory, 
showing that the latter can be represented as general relativity sourced by an effective DM stress energy tensor. When $\mathcal{F}_0\neq 0$, i.e. when a scalar coupling is present, this effective stress energy tensor features both an isotropic pressure and an anisotropic stress of non local nature, even if one started with the assumption of a pressureless perfect fluid for cold DM;

\item we have proven that in the Newtonian limit our theory reduces to the fractional gravity setup of our previous works if the non-local operators satisfy the relation $\mathcal{F}_2(\Box) - 2\,\mathcal{F}_0(\Box)= \Box^{-1} + \ell^2(-\ell^2\Box)^{-s}$;

\item  we have shown that in the Newtonian limit the deviation of fractional gravity with respect to the standard Newtonian setup can be interpreted in terms of a non local isotropic pressure, which gravitates even for a weak and quasi static source. Remarkably, the effective density sourcing the fractional Poisson Equation \eqref{eq|FPE} can be written as $\rho_{\rm DM, eff} = \rho_{\rm DM} + 3\bar p_{\rm DM}$, independently on the particular choice of $\mathcal{F}_0(\Box)$ and $\mathcal{F}_2(\Box)$.

\item we have applied the theory of gravitational lensing to our setup, finding the expression of the convergence in the thin lens approximation. Then we have tested two particularly interesting realizations of our general relativistic theory, namely a pure tensorial model (RTFG) and one where the scalar and tensorial part combine in a coupling involving the Einstein tensor (EFG), against the lensing data of the individual and stacked clusters from the CLASH sample. The general tendency is for FG to perform well on this data sample, at a level comparable with standard general relativity.

\end{itemize}

In future works we plan to: analyze the behavior of the general theory presented here in a standard cosmological setting; look for specific solutions in the strong gravity regime, focusing on ultra-compact objects and black holes; investigate the physical mechanism behind the appearance of the non local coupling between DM and gravity, taking inspiration from the models which exploit fractional calculus to describe anomalous diffusion and brownian motion in fluids.

\section*{Acknowledgments}
We acknowledge the anonymous referee for insightful and constructive comments. We warmly thank G. Gandolfi for useful discussions. This work is partially funded from the projects: ``Data Science methods for MultiMessenger Astrophysics \& Multi-Survey Cosmology'' funded by the Italian Ministry of University and Research, Programmazione triennale 2021/2023 (DM n.2503 dd. 9 December 2019), Programma Congiunto Scuole; Italian Research Center on High Performance Computing Big Data and Quantum Computing (ICSC), project funded by European Union - NextGenerationEU - and National Recovery and Resilience Plan (NRRP) - Mission 4 Component 2 within the activities of Spoke 3 (Astrophysics and Cosmos Observations);  European Union - NextGenerationEU under the PRIN MUR 2022 project n. 20224JR28W "Charting unexplored avenues in Dark Matter"; INAF Large Grant 2022 funding scheme with the project "MeerKAT and LOFAR Team up: a Unique Radio Window on Galaxy/AGN co-Evolution; INAF GO-GTO Normal 2023 funding scheme with the project "Serendipitous H-ATLAS-fields Observations of Radio Extragalactic Sources (SHORES)". CB acknowledges support from the InDark INFN Initiative.

\begin{table*}[h]
\caption{Marginalized posterior estimates (mean and $1\sigma$ confidence intervals are reported) for the parameters from the MCMC analysis of the CLASH clusters in EFG (first lines), in RTFG (second lines) and in general relativity (third lines, with $s=1$). Columns report the values of fractional index $s$, fractional length-scale $\log \ell/r_s$, concentration $\log c_{200}$, DM mass $\log M_{200} [M_\odot]$ in solar units, and the values of the reduced $\chi_r^2$ for the overall fit.}\label{tab|fitparams}
\begin{tabular}{lcccccccccc}
\textbf{Cluster} & & \boldmath{$s$} & &\boldmath{$\log \ell/r_s$} & &\boldmath{$\log c_{200}$}  & & \boldmath{$\log M_{200}$} & & \boldmath{$\chi_r^2$} \\
\hline
\hline
& & $1.28^{+0.15}_{-0.11}$ & & $-0.08^{+0.93}_{-0.30}$ & & $0.77^{+0.14}_{-0.13}$ & & $14.99^{+0.30}_{-0.19}$ & & $1.18$ \\
A383& & $1.11^{+0.05}_{-0.09}$ & & $-0.72^{+1.00}_{-0.60}$ & & $0.68^{+0.11}_{-0.16}$ & & $14.93^{+0.32}_{-0.21}$ & & $1.64$ \\
& & $1$ & & & & $0.44^{+0.10}_{-0.09}$ & & $15.30^{+0.07}_{-0.07}$ & & $1.49$ \\
\hline
& & $<1.14$ & & $>-0.99$ & & $0.57^{+0.09}_{-0.10}$ & & $15.14^{+0.21}_{-0.11}$ & & $0.84$ \\
A209& & $<1.10$ & & $-0.61^{+1.20}_{-0.64}$ & & $0.59^{+0.09}_{-0.12}$ & & $15.09^{+0.24}_{-0.13}$ & & $0.83$ \\
& & $1$ & & & & $0.48^{+0.09}_{-0.09}$ & & $15.32^{+0.08}_{-0.08}$ & & $0.68$ \\
\hline
& & $<1.13$ & & $-0.9^{+1.3}_{-1.0}$ & & $0.70^{+0.08}_{-0.10}$ & & $15.12^{+0.27}_{-0.13}$ & & $0.49$ \\
A2261& & $<1.08$ & & $-1.11^{+0.98}_{-0.98}$ & & $0.69^{+0.08}_{-0.10}$ & & $15.08^{+0.27}_{-0.14}$ & & $0.57$ \\
& & $1$ & & & & $0.63^{+0.08}_{-0.08}$ & & $15.36^{+0.06}_{-0.06}$ & & $0.49$ \\
\hline
& & $<1.18$ & & $>-0.557$ & & $0.72^{+0.09}_{-0.13}$ & & $14.79^{+0.20}_{-0.11}$ & & $0.56$ \\
RXJ2129& & $<1.08$ & & $-0.56^{+1.50}_{-0.49}$ & & $0.70^{+0.09}_{-0.12}$ & & $14.75^{+0.21}_{-0.11}$ & & $0.63$ \\
& & $1$ & & & & $0.62^{+0.09}_{-0.09}$ & & $14.91^{+0.08}_{-0.08}$ & & $0.56$ \\
\hline
& & $<1.15$ & & $>-0.922$ & & $0.67^{+0.09}_{-0.11}$ & & $15.08^{+0.22}_{-0.12}$ & & $0.42$ \\
A611& & $<1.08$ & & $>-0.74^{+1.30}_{-0.79}$ & & $0.66^{+0.09}_{-0.11}$ & & $15.03^{+0.24}_{-0.13}$ & & $0.42$ \\
& & $1$ & & & & $0.56^{+0.09}_{-0.09}$ & & $15.24^{+0.08}_{-0.08}$ & & $0.36$ \\
\hline
& & $1.26^{+0.12}_{-0.12}$ & & $-0.21^{+0.96}_{-0.36}$ & & $0.71^{+0.14}_{-0.14}$ & & $14.94^{+0.30}_{-0.20}$ & & $0.79$ \\
MS2137& & $1.11^{+0.05}_{-0.09}$ & & $-0.68^{+1.00}_{-0.61}$ & & $0.65^{+0.11}_{-0.15}$ & & $14.91^{+0.32}_{-0.20}$ & & $0.99$ \\
 & & $1$ & & & & $0.44^{+0.11}_{-0.10}$ & & $15.28^{+0.09}_{-0.09}$ & & $0.86$ \\
\hline
& & $1.23^{+0.10}_{-0.19}$ & & $>0$ & & $0.63^{+0.12}_{-0.12}$ & & $15.29^{+0.20}_{-0.14}$ & & $0.47$ \\
RXJ2248& & $<1.11$ & & $-0.21^{+1.20}_{-0.27}$ & & $0.61^{+0.11}_{-0.14}$ & & $15.24^{+0.22}_{-0.13}$ & & $0.58$ \\
& & $1$ & & & & $0.41^{+0.12}_{-0.10}$ & & $15.39^{+0.09}_{-0.09}$ & & $0.51$ \\
\hline
& & $<1.17$ & & $>-0.72$ & & $0.58^{+0.09}_{-0.11}$ & & $15.11^{+0.21}_{-0.11}$ & & $0.61$ \\
MACSJ1115& & $<1.12$ & & $>-0.42^{+1.20}_{-0.48}$ & & $0.59^{+0.09}_{-0.13}$ & & $15.07^{+0.23}_{-0.13}$ & & $0.58$ \\
& & $1$ & & & & $0.42^{+0.10}_{-0.10}$ & & $15.25^{+0.07}_{-0.07}$ & & $0.48$ \\
\hline
& & $1.25^{+0.11}_{-0.11}$ & & $-0.55^{+0.91}_{-0.52}$ & & $0.64^{+0.13}_{-0.13}$ & & $15.05^{+0.32}_{-0.26}$ & & $0.99$ \\
MACSJ1931& & $1.16^{+0.08}_{-0.09}$ & & $-0.70^{+0.81}_{-0.43}$ & & $0.63^{+0.13}_{-0.15}$ & & $15.00^{+0.28}_{-0.28}$ & & $1.05$ \\
& & $1$ & & & & $0.34^{+0.12}_{-0.11}$ & & $15.56^{+0.09}_{-0.09}$ & & $0.91$ \\
\hline
& & $1.25^{+0.09}_{-0.11}$ & & $-0.32^{+0.98}_{-0.47}$ & & $0.68^{+0.13}_{-0.13}$ & & $15.01^{+0.32}_{-0.23}$ & & $0.98$ \\
MACSJ1720& & $1.15^{+0.06}_{-0.07}$ & & $-0.57^{+0.77}_{-0.50}$ & & $0.67^{+0.13}_{-0.15}$ & & $14.92^{+0.30}_{-0.25}$ & & $1.05$ \\
& & $1$ & & & & $0.39^{+0.10}_{-0.10}$ & & $15.40^{+0.09}_{-0.09}$ & & $1.05$ \\
\hline
& & $<1.18$ & & $>-0.36$ & & $0.54^{+0.09}_{-0.09}$ & & $15.02^{+0.18}_{-0.12}$ & & $1.03$ \\
MACSJ0416& & $1.09^{+0.04}_{-0.08}$ & & $-0.09^{+1.01}_{-0.22}$ & & $0.58^{+0.11}_{-0.13}$ & & $14.97^{+0.20}_{-0.13}$ & & $0.99$ \\
& & $1$ & & & & $0.36^{+0.09}_{-0.09}$ & & $15.09^{+0.07}_{-0.08}$ & & $0.84$ \\
\hline
& & $1.19^{+0.08}_{-0.15}$ & & $-0.36^{+1.20}_{-0.39}$ & & $0.66^{+0.12}_{-0.12}$ & & $14.92^{+0.26}_{-0.18}$ & & $0.55$ \\
MACSJ0429& & $1.11^{+0.04}_{-0.09}$ & & $-0.51^{+1.00}_{-0.54}$ & & $0.66^{+0.12}_{-0.14}$ & & $14.87^{+0.28}_{-0.19}$ & & $0.57$ \\
& & $1$ & & & & $0.44^{+0.13}_{-0.11}$ & & $15.19^{+0.09}_{-0.10}$ & & $0.54$ \\
\hline
& & $1.24^{+0.12}_{-0.14}$ & & $>-0.03$ & & $0.68^{+0.12}_{-0.12}$ & & $15.19^{+0.23}_{-0.12}$ & & $0.72$ \\
MACSJ1206& & $1.09^{+0.04}_{-0.08}$ & & $-0.32^{+1.10}_{-0.44}$ & & $0.64^{+0.11}_{-0.15}$ & & $15.13^{+0.25}_{-0.14}$ & & $0.91$ \\
& & $1$ & & & & $0.43^{+0.09}_{-0.09}$ & & $15.34^{+0.07}_{-0.07}$ & & $0.85$ \\
\hline
& & $<1.11$ & & $>-0.98$ & & $0.72^{+0.09}_{-0.11}$ & & $14.92^{+0.18}_{-0.09}$ & & $0.97$ \\
MACSJ0329& & $<1.07$ & & $-0.70^{+1.60}_{-0.58}$ & & $0.72^{+0.09}_{-0.11}$ & & $14.88^{+0.20}_{-0.11}$ & & $0.98$ \\
& & $1$ & & & & $0.66^{+0.11}_{-0.11}$ & & $15.02^{+0.08}_{-0.08}$ & & $0.83$ \\
\hline
& & $<1.15$ & & $>-0.72$ & & $0.60^{+0.10}_{-0.10}$ & & $15.41^{+0.18}_{-0.11}$ & & $0.89$ \\
RXJ1347& & $<1.08$ & & $-0.56^{+1.5}_{-0.45}$ & & $0.61^{+0.09}_{-0.12}$ & & $15.37^{+0.20}_{-0.12}$ & & $0.89$ \\
& & $1$ & & & & $0.49^{+0.11}_{-0.11}$ & & $15.51^{+0.07}_{-0.08}$ & & $0.76$ \\
\hline
& & $1.31^{+0.10}_{-0.09}$ & & $0.08^{+0.58}_{-0.35}$ & & $0.58^{+0.12}_{-0.12}$ & & $15.14^{+0.28}_{-0.20}$ & & $0.85$ \\
MACSJ1149& & $1.20^{+0.07}_{-0.07}$ & & $0.01^{+0.45}_{-0.37}$ & & $0.63^{+0.14}_{-0.14}$ & & $15.08^{+0.28}_{-0.24}$ & & $0.86$ \\
& & $1$ & & & & $0.17^{+0.09}_{-0.09}$ & & $15.41^{+0.07}_{-0.07}$ & & $0.83$ \\
\hline
& & $1.17^{+0.07}_{-0.11}$ & & $>-0.13$ & & $0.45^{+0.09}_{-0.09}$ & & $15.29^{+0.18}_{-0.12}$ & & $1.68$ \\
MACSJ0717& & $1.13^{+0.05}_{-0.06}$ & & $+0.23^{+0.64}_{-0.27}$ & & $0.56^{+0.12}_{-0.12}$ & & $15.28^{+0.21}_{-0.14}$ & & $1.49$ \\
& & $1$ & & & & $0.24^{+0.09}_{-0.08}$ & & $15.35^{+0.07}_{-0.07}$ & & $1.38$ \\
\hline
& & $<1.17$ & & $>-0.19$ & & $0.62^{+0.11}_{-0.11}$ & & $15.08^{+0.14}_{-0.11}$ & & $0.68$ \\
MACSJ0647& & $<1.08$ & & $>-0.46$ & & $0.63^{+0.10}_{-0.12}$ & & $15.05^{+0.15}_{-0.11}$ & & $0.68$ \\
& & $1$ & & & & $0.50^{+0.12}_{-0.11}$ & & $15.11^{+0.09}_{-0.09}$ & & $0.57$ \\
\hline
& & $1.36^{+0.13}_{-0.05}$ & & $0.36^{+0.52}_{-0.21}$ & & $0.68^{+0.11}_{-0.11}$ & & $15.05^{+0.25}_{-0.16}$ & & $1.10$ \\
MACSJ0744& & $1.19^{+0.07}_{-0.07}$ & & $0.09^{+0.47}_{-0.37}$ & & $0.69^{+0.14}_{-0.14}$ & & $14.95^{+0.28}_{-0.2}$ & & $1.42$ \\
& & $1$ & & & & $0.23^{+0.11}_{-0.11}$ & & $15.27^{+0.08}_{-0.08}$ & & $1.45$ \\
\hline
& & $1.18^{+0.07}_{-0.08}$ & & $>0.28$ & & $0.64^{+0.09}_{-0.06}$ & & $14.88^{+0.14}_{-0.06}$ & & $0.74$ \\
STACKED& & $1.10^{+0.04}_{-0.04}$ & & $>0.14$ & & $0.69^{+0.11}_{-0.11}$ & & $14.84^{+0.16}_{-0.07}$ & & $0.83$ \\
& & $1$ & &  & & $0.46^{+0.03}_{-0.03}$ & & $14.91^{+0.03}_{-0.02}$ & & $0.94$ \\
\hline
\\ 
\end{tabular}
\end{table*}

\bibliography{biblio}
\bibliographystyle{unsrt}

\end{document}